\documentclass[aps,prb,twocolumn,showpacs]{revtex4}
\usepackage{amsfonts}
\usepackage{subfigure}
\usepackage{amssymb}
\usepackage{amsmath}
\usepackage{bm}
\usepackage[dvips]{graphicx}
\usepackage[dvips]{color}
\usepackage{graphicx}
\usepackage{epsfig}
\usepackage{hyperref}

\begin{document}

\title{Spin Qubit Relaxation in a Moving Quantum Dot}
\author{Peihao Huang}
\email{peihaohu@buffalo.edu}
\author{Xuedong Hu}
\affiliation{Department of Physics, University at Buffalo, SUNY, Buffalo, NY 14260, USA}

\begin{abstract}
Long-range quantum communication for spin qubits is an important open
problem. Here we study decoherence of an electron spin qubit that is being
transported in a moving quantum dot. We focus on spin decoherence due to
spin-orbit interaction and a random electric potential. We find that at the
lowest order, the motion induces longitudinal spin relaxation, with a rate
linear in the dot velocity. Our calculated spin relaxation time ranges from
sub $\mu$s in GaAs to above ms in Si, making this relaxation a significant
decoherence channel. Our results also give clear indications on how to
reduce the decoherence effect of electron motion.
\end{abstract}

\date{\today}
\maketitle

%\email{xhu@buffalo.edu}

\section{introduction}

Over the past decade, there has been tremendous progress in the
experimental study and theoretical investigation of spin qubit manipulation
and decoherence \cite{Hanson2007,Morton2011}. An important advantage of
electron spin qubits is that they can be coupled strongly via the exchange
interaction \cite{Loss1998}, which allows fast two-spin gates. However,
exchange interaction is short-ranged, and long-range quantum communication
remains a significant open problem in the scale-up of spin qubit
architectures.

Various ideas have been proposed to move spin information on chip, such as
via spin-photon coupling \cite{Imamoglu1999,Hu2012}, spin bus \cite%
{Friesen2007}, and directly moving the electrons themselves \cite%
{Barnes2000,Skinner2003, Greentree2004, Taylor2005}. Spin-photon conversion
is difficult because the strongest estimated coupling strength is still only
in the order of MHz \cite{Hu2012, Petersson2012}, which is slower than or
comparable to the spin dephasing rate in most materials. Spin buses, on the
other hand, are limited by the energy gap between the bus ground state(s)
and the excited states \cite{Friesen2007, Oh2011}, and are most useful for
short-distance (up to a few microns) quantum information transfer.
Comparatively, direct spin transport is attractive because of its conceptual
simplicity and its similarity to the conventional charge-coupled devices.
Indeed, several experimental groups have recently shown how a surface
acoustic wave (SAW) can controllably transport an electron over several
microns \cite{Stotz2005, Hermelin2011, McNeil2011, Sanada2011, Yamamoto2012}%
. Clearly, transferring spin information by directly moving its carrier is
an intriguing and promising approach, and deserves further in-depth
analysis. Here we focus on the aspect of spin decoherence due to electron
motion.

The main decoherence channel for a confined electron spin in a finite field
is the hyperfine (HF) interaction induced pure dephasing \cite{Cywinski2009,
Bluhm2011}. Spin relaxation due to spin-orbit (SO) interaction is much
slower \cite{Golovach2004}. On the other hand, spin relaxation of free
electrons and holes in semiconductors is dominated by SO interaction \cite%
{Meier1984,Zutic2004}, while the effect of hyperfine interaction
is strongly suppressed by motional narrowing \cite{Zutic2004}. For a moving
electron spin qubit with controlled motion, an intriguing question is thus
when decoherence due to SO interaction becomes dominant.

In this paper we study spin decoherence of a moving but confined electron
due to static disorders in a semiconductor heterostructure. For example, in
a modulation-doped GaAs/Al$_x$Ga$_{1-x}$As structure, the ionized dopants
produce a random electric potential at the GaAs interface where the quantum
dot (QD) is located. If the QD is moved
%(by programming the top gate potentials)
along the interface, the electron spin can sense this random potential
through the SO interaction, and undergo decoherence. The static disorder has been considered in the problem of spin relaxation of 2D electrons \cite{Dugaev2009,Glazov2010}, while we focus on its effect on confined (albeit moving) electrons. Here we construct a
theoretical description of this decoherence mechanism. We find that this is
a longitudinal relaxation channel at the lowest order. Its rate can be as
fast as sub-$\mu$s in GaAs or above ms in Si, making it a significant
decoherence channel.

%\textit{Electron Hamiltonian in the Moving Frame---}

\section{Electron Hamiltonian}

The model system we consider is a single electron in a gate-defined QD from
a two-dimensional electron gas (2DEG). In general the growth-direction
confinement is much stronger, so that the QD can only be moved in the
in-plane direction by programming the top gate potentials. The electron
remains confined while the QD is moved. Indeed, we assume the QD motion is
adiabatic so that the electron remains in the ground orbital state.

\begin{figure}[ht]
% Requires \usepackage{graphicx}
\centering
\includegraphics[scale=0.75]{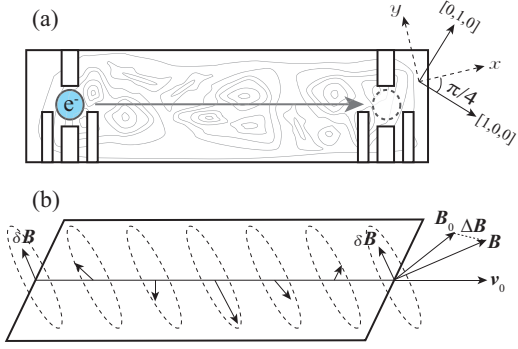}
\caption{A schematic of a spin qubit in a moving QD. Panel (a) gives the
topview of the structure and the coordinate system ($xyz$) defined in the
laboratory frame, with $x$ and $y$ along the [$110$] and [$\bar{1}10$]
directions. Panel (b) gives the sideview and the total effective magnetic
field.}
\label{Random Potential}
\end{figure}

As shown in Fig.~\ref{Random Potential}, we consider a uniform linear motion
of the QD with a constant velocity $\boldsymbol{v}_{0}$ (the QD potential
minimum is at $\boldsymbol{r}_{0}\left( t\right) =\boldsymbol{v}_{0}t=\left[
x_{0}\left( t\right) ,y_{0}\left( t\right) \right] $). Such a linear motion is possible in principle by programming the surface gate potential along the path of QD motion.\cite{Skinner2003,Taylor2005}  Alternatively, SAW has been shown to be effective in moving a single electron from one QD to another at the speed of sound.\cite{Barnes2000,Stotz2005,Hermelin2011,McNeil2011,Sanada2011,Yamamoto2012}.  In both these cases, the QD motion is facilitated by an external agent (external electrical control in the former, and the piezoelectric generator of the SAW in the latter).  As such, the complete problem of the electron dynamics in a moving quantum dot is that for an open system, where the reservoir (or the external driving agent) is complex and difficult to define completely and quantum mechanically. In the present study, in order to simplify the problem, we assume that the reservoir, or the external driving agent, is classical and Markovian. In other words, the driving agent is so large that it does not remember any energy exchange with the electron. The consequences of this assumption will be discussed later in the manuscript. The Hamiltonian for the QD-confined electron is
\begin{eqnarray}
H &=&H_{d}+H_{Z}+H_{SO}+\delta V\left( \boldsymbol{r}\right) ,  \label{H} \\
H_{d} &=&\frac{\boldsymbol{\pi }^{2}}{2m^{\ast }}+U\left( \boldsymbol{r}-%
\boldsymbol{r}_{0}\left( t\right) \right) , \\
H_{Z} &=&\frac{1}{2}g\mu _{B}{\boldsymbol{B}}_{0}\cdot {\boldsymbol{\sigma }}%
, \\
H_{SO} &=&\beta _{-}\pi _{y}\sigma _{x}+\beta _{+}\pi _{x}\sigma _{y},
\label{Hso}
\end{eqnarray}%
where $\delta V\left( \boldsymbol{r}\right) $ represents a random electric
potential, which is always present, whether due to modulation doping or
barrier disorder. The subscripts $d$, $Z$, and $SO$ refer to "dot",
"Zeeman", and "spin-orbit". In $H_{d}$, ${\boldsymbol{\pi }}$ is the
electron 2D momentum ($e>0$), given by $\boldsymbol{\pi }=-i\hbar {%
\boldsymbol{\nabla }}+(e/c)\boldsymbol{A}(\boldsymbol{r})$, and $U\left(
\boldsymbol{r}-\boldsymbol{r}_{0}\left( t\right) \right) $ is the dot
confinement potential with a moving minimum $\boldsymbol{r}_{0}\left(
t\right) =\left[ x_{0}\left( t\right) ,y_{0}\left( t\right) \right] $. In
this study, we consider a uniform linear motion of the QD, where $%
\boldsymbol{v}_{0}=\boldsymbol{p}_{0}/m^{\ast }=d\boldsymbol{r}_{0}\left(
t\right) /dt$ is a constant vector. In $H_{Z}$, ${\boldsymbol{B}}_{0}$ is
the applied magnetic field (with $\boldsymbol{\hat{n}}_{0}$ its unit
vector). In $H_{SO}$, $\beta _{\pm }\equiv \left( \beta \pm \alpha \right) $%
, where $\alpha $ and $\beta $ are the Rashba and Dresselhaus SO coupling
constants. The $x$ and $y$ axes are along the [$110$] and [$\bar{1}10$]
directions. If $x$ and $y$ had been defined along the [$100$] and [$010$]
directions, the SO term would have taken the usual form $H_{SO}=\beta (-\pi
_{x}\sigma _{x}+\pi _{y}\sigma _{y})+\alpha (\pi _{x}\sigma _{y}-\pi
_{y}\sigma _{x})$. The current choice of $x$ and $y$ helps simplify the
presentation below.

To simplify the following treatment, we transform into the moving reference
frame, so that the QD confinement potential becomes time-independent. It is
done by a translational transformation $\left\vert \psi ^{\prime }\left(
t\right) \right\rangle =\exp \left[ S_{T}\left( t\right) \right] \left\vert
\psi \left( t\right) \right\rangle $, where the generator is $S_{T}\left(
t\right) \equiv i\boldsymbol{\pi }\cdot \boldsymbol{r}_{0}\left( t\right)
/\hbar $. The Schr\"{o}dinger equation after the transformation is%
\begin{equation}
i\hbar \frac{\partial }{\partial t}\left\vert \psi ^{\prime }\left( t\right)
\right\rangle =H^{\prime }\left\vert \psi ^{\prime }\left( t\right)
\right\rangle \,,
\end{equation}%
in which, the new Hamiltonian is
$
H^{\prime }=e^{S_{T}}He^{-S_{T}}+i\hbar \partial _{t}S_{T}.
$
After the transformation, the total Hamiltonian in
the moving frame is
\begin{eqnarray}
H^{\prime }&=&H_{d}^{\prime }+H_{Z}^{\prime }+H_{SO}^{\prime }+\delta V\left(
\boldsymbol{r}_{0}\left( t\right) +\boldsymbol{r}^{\prime }\right) ,\label{Hp} \\
H_{d}^{\prime } &=&\frac{\boldsymbol{\pi }^{\prime 2}}{2m^{\ast }}+U\left(
\boldsymbol{r}^{\prime }\right) , \\
H_{Z}^{\prime } &=&\frac{1}{2}g\mu _{B}{\boldsymbol{B}}\cdot {\boldsymbol{%
\sigma }}, \\
H_{SO}^{\prime } &=&\beta _{-}\pi _{y^{\prime }}^{\prime }\sigma _{x^{\prime
}}+\beta _{+}\pi _{x^{\prime }}^{\prime }\sigma _{y^{\prime }}.  \label{Hsop}
\end{eqnarray}%
Here $\boldsymbol{r^{\prime }}=\left[
x^{\prime },y^{\prime }\right] $ and $\boldsymbol{%
\pi ^{{\boldsymbol{\prime }}}}$ are the electron two-dimensional position
and momentum operators in the moving reference frame. Operators
in different frames are related: $\boldsymbol{r}=\boldsymbol{r^{\prime }}+%
\boldsymbol{r}_{0}\left( t\right) $ with $\boldsymbol{r}_{0}\left( t\right) =%
\boldsymbol{v}_{0}t$ and $\boldsymbol{\pi ^{{\boldsymbol{\prime }}}}\equiv -i\hbar {\boldsymbol{%
\nabla }}\boldsymbol{^{{\boldsymbol{\prime }}}}+(e/c)\boldsymbol{A}(%
\boldsymbol{r^{{\boldsymbol{\prime }}}})-\boldsymbol{p}_{L}-\boldsymbol{p}%
_{0}$, in which $\boldsymbol{p}_{L}\left( t\right)
\equiv -eB_{0z}/c\left[ -y_{0}\left( t\right) ,x_{0}\left( t\right) ,0\right]
$ captures the effect of the Lorentz force (note that the classical motion
of electron satisfies $\frac{d\boldsymbol{p}}{dt}%
=-e/c\boldsymbol{v}_{0}\times \boldsymbol{B}_{0} =\frac{d\boldsymbol{p}_{L}%
}{dt}$). In the moving frame, the random
potential is time-dependent due to the QD motion, $\delta V=\delta V\left(
\boldsymbol{r}_{0}\left( t\right) +\boldsymbol{r}^{\prime }\right) $, so
that the static disorder is now a charge noise. %
%, {\it i.e.} the \textit{%
%spatially} random potential in the lab frame is \textit{temporally} random in the moving frame.
%
In $H_{d}$, $U\left( \boldsymbol{r}^{\prime }\right)$ is the time independent confinement
potential in the moving frame. In $H_{Z}$, ${\boldsymbol{B}}={\boldsymbol{B}}%
_{0}+\Delta {\boldsymbol{B}}$ is the total magnetic field, in which $\Delta {%
\boldsymbol{B}}$ is an effective magnetic field due to SO interaction in the
moving reference frame
\begin{equation}
\Delta {\boldsymbol{B}}=\frac{2m^{\ast }v_{0}}{g\mu _{B}}\left( \beta
_{-}\sin \phi _{v},\beta _{+}\cos \phi _{v},0\right) ,  \label{DeltaB}
\end{equation}%
where $\phi _{v}$ is the angle between the dot velocity and the [$110$]
crystal axis. The Zeeman frequency is $\omega _{Z}=g\mu _{B}B/\hbar $ and
the spin quantization direction is $\boldsymbol{\hat{n}}={\boldsymbol{B/}}B$%
, which is generally different from the direction of the applied magnetic
field ${\boldsymbol{B}}_{0}$.

\section{Constructing the Effective Spin Hamiltonian for a moving electron}

As the quantum dot is moved in a semiconductor heterostructure, the
spatially random electrical potential $\delta V\left( \boldsymbol{r}_{0}+%
\boldsymbol{r}^{\prime }\right) $ causes the momentum of the QD-confined
electron to fluctuate. The electron spin can sense these momentum
fluctuations via the spin-orbit Hamiltonian (\ref{Hsop}), and spin
decoherence ensues. The QD motion we consider here is sufficiently slow so
that it does not lead to any orbital excitation. We can focus on the
electron spin dynamics by decoupling the spin space (with the ground orbital
state) from the rest of the Hilbert space \cite%
{Golovach2004,Borhani2006,Golovach2006,Aleiner2001,Stano2006}. Specifically,
we perform a Schrieffer-Wolff transformation $\tilde{H}=e^{S}He^{-S}$ to
remove the SO coupling in the leading order by requiring that $\left[
H_{d}^{\prime }+H_{Z}^{\prime },S\right] =H_{SO}^{\prime }$. \cite%
{Golovach2004,Borhani2006,Golovach2006}.
For the harmonic confinement $%
U(r^{\prime })=\frac{1}{2}m^{\ast }\omega _{d}^{2}r^{\prime 2}$, the SO term
can be expressed as $H_{SO}^{\prime }=i\mathbb{L}_{d}\left( \boldsymbol{%
\sigma }\cdot \boldsymbol{\xi }\right) $, where $\mathbb{L}%
_{d}A\equiv \lbrack H_{d}^{\prime },A]$, $\forall A$ and ${\boldsymbol{\xi }}\equiv
(y^{\prime }/\lambda _{-},x^{\prime }/\lambda _{+},0)$ is a vector in the
2DEG plane; $\lambda _{\pm }\equiv \hbar /\left( m^{\ast }\beta
_{\pm }\right) $ are the spin-orbit lengths. The superoperator $\mathbb{L}%
_{d}$ satisfies $\mathbb{L}_{d}^{-1}\boldsymbol{\pi }^{\prime }=im^{\ast }\boldsymbol{r}%
^{\prime }/\hbar $, and $\mathbb{L}_{d}^{-1}\left[ x^{\prime },y^{\prime }%
\right] =-i\left( \hbar m^{\ast }\omega _{d}^{2}\right) ^{-1}\left[ \left(
\pi _{x^{\prime }}^{\prime }+m^{\ast }\omega _{c}y^{\prime }\right) ,\left(
\pi _{y^{\prime }}^{\prime }-m^{\ast }\omega _{c}x^{\prime }\right) \right] $%
, where $E_{Z}=g\mu _{B}B$ is the electron Zeeman splitting (with $\omega
_{Z}\equiv E_{Z}/\hbar $ being the Zeeman frequency) and $\omega _{c}\equiv
eB_{0z}/\left( m^{\ast }c\right) $ is the cyclotron frequency. Assuming that
the Zeeman energy is much larger than the SO energy, but much smaller than
the orbital excitation energy ($m^{\ast }\left( \beta ^{2}+\alpha
^{2}\right) \ll \hbar \omega _{Z}\ll \hbar \omega _{d}$), we get \cite%
{Golovach2004,Borhani2006,Golovach2006},
\begin{equation}
S=i\boldsymbol{\sigma }\cdot \boldsymbol{\xi +}E_{Z}\left( \hat{\boldsymbol{n%
}}\times \mathbb{L}_{d}^{-1}\boldsymbol{\xi }\right) \cdot \boldsymbol{%
\sigma }.
\end{equation}%
The transformed Hamiltonian is thus %
\begin{equation}
H^{\prime \prime }=i\hbar \partial _{t}S+H_{d}^{\prime }+H_{Z}^{\prime }+
\left[ S,\delta V\left( \boldsymbol{r}\right) \right] +\cdots ,
\end{equation}%
in which $i\hbar \partial _{t}S=\frac{1}{2}g\mu _{B}\Delta \boldsymbol{B}%
_{L}\cdot \boldsymbol{\sigma }$ with
$%\begin{equation}
\Delta \boldsymbol{B}_{L}=\frac{\omega _{Z}\omega _{c}}{\omega _{d}^{2}}\hat{%
\boldsymbol{n}}\times \frac{2\hbar }{g\mu _{B}}\left[ v_{0x}/\lambda
_{-},-v_{0y}/\lambda _{+},0\right]  \label{DeltaB_L}
$%\end{equation}%
being the high order correction of $\Delta {\boldsymbol{B}}$ which will be dropped in the following
discussion.
%, thus $\Delta \boldsymbol{%
%B}_{L}=\frac{\omega _{Z}\omega _{c}}{\omega _{d}^{2}}\hat{\boldsymbol{n}}%
%\times \left[ \lambda _{+}\Delta B_{y}/\lambda _{-},-\lambda _{-}\Delta
%B_{x}/\lambda _{+},0\right] $. As shown below, $\Delta B\ll 1$ T.
%Furthermore, $\omega _{Z}\ll \omega _{d}$ and $\omega _{c}\sim 1$ meV $\sim
%\omega _{d}$ for a perpendicular 1 T magnetic field in GaAs ($\omega _{c}=0$
%for the in-plane magnetic field). Therefore, the correction $\Delta
%\boldsymbol{B}_{L}$ due to the Lorentz force is generally much smaller than $%
%\Delta {\boldsymbol{B}}$, and will be dropped from the following
%considerations.

The first order term due to the random electric potential is $\left[
S,\delta V\left( \boldsymbol{r}\right) \right] =\frac{1}{2}g\mu _{B}2%
\boldsymbol{B}\times \boldsymbol{\Omega }\left( \boldsymbol{r}\right) \cdot
\boldsymbol{\sigma }$, in which
\begin{equation}
\boldsymbol{\Omega }\left( \boldsymbol{r}\right) =\frac{-e}{m^{\ast }\omega
_{d}^{2}}\left[ \varepsilon _{y^{\prime }}\left( \boldsymbol{r}\right)
/\lambda _{-},\varepsilon _{x^{\prime }}\left( \boldsymbol{r}\right)
/\lambda _{+},0\right] ,
\end{equation}%
where the electric field corresponding to the random potential is $%
\boldsymbol{\varepsilon }\left( \boldsymbol{r}\right) =\frac{1}{e}\nabla
\delta V\left( \boldsymbol{r}\right) $ ($e>0$). Therefore, the effective
spin Hamiltonian takes the form
\begin{eqnarray}
H_{eff} &=&\frac{1}{2}g\mu _{B}(\boldsymbol{B}+\delta \boldsymbol{B}%
(t))\cdot \boldsymbol{\sigma },  \label{Heff} \\
\delta \boldsymbol{B}(t) &=&2\boldsymbol{B}\times \boldsymbol{\Omega }\left(
t\right) ,  \label{dB} \\
\boldsymbol{\Omega }\left( t\right)  &=&\left\langle \psi \right\vert
\boldsymbol{\Omega }\left( \boldsymbol{r}_{0}\left( t\right) +\boldsymbol{%
r^{\prime }}\right) \left\vert \psi \right\rangle ,  \label{Omega_r}
\end{eqnarray}%
where $\left\vert \psi \right\rangle $ is the orbital wave function.
To further simplify Eq. (\ref{Omega_r}), we expand the random
electric field $\varepsilon _{i}\left( \boldsymbol{r}\right) $ ($i=x^{\prime
}$, $y^{\prime }$) around the average QD position $\boldsymbol{r}_{0}$,
\begin{equation*}
\varepsilon _{i}\left( \boldsymbol{r}\right) \approx \varepsilon _{i}\left(
\boldsymbol{r}_{0}\right) +\boldsymbol{\nabla }\varepsilon _{i}\left(
\boldsymbol{r}_{0}\right) \cdot \boldsymbol{r^{\prime }}+\frac{1}{2!}%
\boldsymbol{\nabla \nabla }\varepsilon _{i}\left( \boldsymbol{r}_{0}\right)
\cdot \boldsymbol{r^{\prime }\cdot r^{\prime }}+\cdots .
\end{equation*}%
Due to the adiabatic condition, the electron always remains in the
instantaneous ground orbital state $\psi (\boldsymbol{r}^{\prime })=\exp
\left( -r^{\prime 2}/2\lambda ^{2}\right) /\lambda \sqrt{\pi }$, up to a
magnetic phase that does not affect the calculation of $\Omega $. Here $%
\lambda ^{-2}=\hbar ^{-1}\sqrt{(m^{\ast }\omega _{d})^{2}+(eB_{z}/2c)^{2}}$.
For small variation of the gradient of the electric field $\lambda ^{2}%
\boldsymbol{\nabla \nabla }\varepsilon _{i}\left( \boldsymbol{r}_{0}\right)
\ll \varepsilon _{i}\left( \boldsymbol{r}_{0}\right) $ (and keep in mind
that the average of the linear term with a symmetric ground state wave
function vanishes), we retain only the zeroth order of $\varepsilon
_{i}\left( \boldsymbol{r}\right) $, so that
\begin{equation}
\boldsymbol{\Omega }\left( t\right) =\frac{-e}{m^{\ast }\omega _{d}^{2}}%
\left[ \varepsilon _{y^{\prime }}\left( \boldsymbol{r}_{0}\left( t\right)
\right) /\lambda _{-},\varepsilon _{x^{\prime }}\left( \boldsymbol{r}%
_{0}\left( t\right) \right) /\lambda _{+},0\right] .  \label{Omega}
\end{equation}%

The effective Hamiltonian holds in the lowest order of spin-orbit interaction
and the lowest order of Zeeman splitting ($\Delta \boldsymbol{B}_{L}$ goes
to zero in this limit), and it has two important features. First, the \emph{%
spatially random} electric field %$\delta {V}(\boldsymbol{r})$ and
$\boldsymbol{\varepsilon }(\boldsymbol{r})$ is now a \emph{temporally random}
magnetic field for the electron spin, ${\delta }\boldsymbol{B}\left(
t\right) $. This transformation is through the dot motion $\boldsymbol{r}%
_{0}\left( t\right) $ [$\boldsymbol{\varepsilon }(\boldsymbol{r})\rightarrow
\boldsymbol{\varepsilon }(t)$] and the SO interaction [$\boldsymbol{%
\varepsilon }(t)\rightarrow \delta \boldsymbol{B}(t)$]. Second, Eqs.~(%
\ref{dB}) shows that there can be \emph{only transverse fluctuations} in the
effective magnetic field since $\delta \boldsymbol{B}(t)\cdot \boldsymbol{B}%
=0$ [see Fig.~\ref{Random Potential}(b)]. Due to this transverse nature,
there is \emph{no pure dephasing} at the lowest order approximation.

%\textit{Noise Correlation---}

\section{Noise Correlation}

To calculate the spin relaxation rate of the moving electron, we need to
first obtain the temporal correlation functions $J_{ij}(t)=\left\langle
\delta B_{i}\delta B_{j}\left( t\right) \right\rangle $ of the random
magnetic field that leads to the spin decoherence. Recall that the random
magnetic field, given by Eqs. (\ref{dB}) and (\ref{Omega}), originates
from the spin-orbit interaction and a random electric field, the latter from
disorder in the substrate material. Thus the temporal correlation of the
magnetic fluctuations comes from the spatial correlation of the random
electric field. Here we choose an isotropic model for the random electrical
field
\begin{equation}
\left\langle \varepsilon _{i}\left( \boldsymbol{r}_{1}\right) \varepsilon
_{j}\left( \boldsymbol{r}_{2}\right) \right\rangle =\delta _{ij}\sigma
_{\varepsilon }^{2}f_{c}\left( \Delta r/l_{\varepsilon }\right) ,
\end{equation}%
where $\delta _{ij}$ is the Kronecker delta function ($i,j=x^{\prime }$ or $%
y^{\prime }$), $\sigma _{\varepsilon }$ is the standard deviation of the
electric field, $f_{c}\left( \Delta r/l_{\varepsilon }\right) $ is the
cutoff function as listed in Table I, in which $\Delta r=\left\vert
\boldsymbol{r}_{1}-\boldsymbol{r}_{2}\right\vert $ is the distance between $%
\boldsymbol{r}_{1}$ and $\boldsymbol{r}_{2}$, and $l_{\varepsilon }$ is the
correlation length of the random field. Here, the average $\left\langle \cdots \right\rangle $ could be the average among different segments of the path when the experiment is done in a straight line, or the average in different directions when the electron is moved that way. It could also be the result of uncertainty in the driving agent, so that the time and the position of the electron do not have a one-to-one correspondence, and this variation in the electron position at a particular time gives us a degree of freedom for the averaging among different paths. Thus in the moving frame
\begin{equation}
\left\langle \varepsilon _{i}\left( \boldsymbol{r}_{0}\left( t_{1}\right)
\right) \varepsilon _{j}\left( \boldsymbol{r}_{0}\left( t_{2}\right) \right)
\right\rangle =\delta _{ij}\sigma _{\varepsilon }^{2}f_{c}\left( \left\vert
t\right\vert /\tau _{c}\right) ,
\end{equation}%
where $t=t_{2}-t_{1}$ and $\tau _{c}=l_{\varepsilon }/v_{0}$ is the
correlation time. The spatially random electric field in the laboratory
frame is now a temporally random electric field in the moving frame, which
through Eqs.~({\ref{dB}}) and (\ref{Omega}) becomes a temporally random
magnetic field.

It is important to emphasize here that we assume the QD trajectories cannot be reproduced identically over different runs, since in realistic situations one cannot drag the electron exactly along the same path.  Most importantly, the moving electron is an open system, constantly exchanging energy with the external driving agent that allows the linear motion of the QD.  Since the external driver is assumed to be classical and Markovian, the electron dynamics is dissipative instead of unitary. Therefore, the resulting spin flip is not a predictable unitary spin rotation, and information is lost due to the exchange with the external reservoir.

\begin{table}[t]
\centering%
\begin{tabular}{|c|c|c|c|}
\hline
& \multicolumn{1}{|c|}{$f_{c}\left( r/l_{\varepsilon }\right) $} & $%
f_{c}\left( \left\vert t\right\vert /\tau _{c}\right) $ & $F_{c}\left(
\omega \right) $ \\ \hline
$1$ & \multicolumn{1}{|r|}{$\exp \left( -r/l_{\varepsilon }\right) $} &
\multicolumn{1}{|r|}{$\exp \left( -\left\vert t\right\vert /\tau _{c}\right)
$} & \multicolumn{1}{|r|}{$2\tau _{c}/\left[ 1+\omega ^{2}\tau _{c}^{2}%
\right] $} \\ \hline
$2$ & \multicolumn{1}{|r|}{$1/\left[ 1+r^{2}/l_{\varepsilon }^{2}\right] $}
& \multicolumn{1}{|r|}{$1/\left[ 1+t^{2}/\tau _{c}^{2}\right] $} &
\multicolumn{1}{|r|}{$\pi \tau _{c}\exp \left( -\left\vert \omega
\right\vert \tau _{c}\right) $} \\ \hline
$3$ & \multicolumn{1}{|r|}{$\exp \left( -r^{2}/l_{\varepsilon }^{2}\right) $}
& \multicolumn{1}{|r|}{$\exp \left( -t^{2}/\tau _{c}^{2}\right) $} &
\multicolumn{1}{|r|}{$\sqrt{\pi }\tau _{c}\exp \left( -\omega ^{2}\tau
_{c}^{2}/4\right) $} \\ \hline
\end{tabular}%
\caption{Fourier transformation of different correlations.}
\label{table1}
\end{table}

The cross product in Eq.~(\ref{dB}) and the arbitrary direction for the
applied magnetic field mean that the magnetic correlation is in general
quite complex in the $(x^{\prime }y^{\prime }z)$ coordinate system we have
used so far. To simplify the relaxation rate calculations, we first
transform to a new $XYZ$ coordinate system, in which we require that (a) $Z$
is along the direction of the total magnetic field $\boldsymbol{B}$ and (b) $%
J_{ij}(t)$ is diagonal in this coordinate system. The first requirement
dictates that $\delta \boldsymbol{B}$ is always in the $XY$ plane since $%
\delta \boldsymbol{B}\perp \boldsymbol{B}$. This means that $\delta B_{Z}=0$
and $J_{ZZ}=0$. The second requirement further dictates that the correlation
functions are diagonal, so that there are only two independent correlation
functions $J_{XX}$ and $J_{YY}$.

The $XYZ$ coordinate system can be obtained from the $(x^{\prime }y^{\prime
}z)$ coordinates by a rotation with Euler angles $\varphi $, $\theta $, and $%
\chi $. Specifically, first rotate $(x^{\prime }y^{\prime }z)$ along the $z$
axis by angle $\varphi$ to $(x^{\prime \prime }y^{\prime \prime }z)$, so
that the $y^{\prime \prime } $ axis is perpendicular to the direction of
magnetic field $\boldsymbol{\hat{n}}$. Then rotate along $y^{\prime \prime }$
by angle $\theta $ to $(x^{\prime \prime \prime }y^{\prime \prime \prime }Z)$%
, so that the $Z$ axis is in the direction $\boldsymbol{\hat{n}}$. Lastly,
rotate along the $Z$ axis by angle $\chi $. Here angles $\varphi $ and $\theta $
give the direction of the total magnetic field $\boldsymbol{B}$ in the $%
(x^{\prime }y^{\prime }z)$ frame, and $\chi$ is determined from the
requirement $\langle \delta B_{X}(0)\delta B_{Y}(t)\rangle =0$.

After the Euler rotations $\boldsymbol{R}_{Z}\left( \chi \right) \boldsymbol{%
R}_{y^{\prime \prime }}\left( \theta \right) \boldsymbol{R}_{z}\left(
\varphi \right) $, the field in the $XYZ$ coordinates is given by $\delta
\boldsymbol{B}(t)=\frac{-2eB}{m^{\ast }\omega _{d}^{2}}\boldsymbol{\zeta }%
\left( t\right) $,
\begin{eqnarray*}
\zeta _{X} &=&\cos \chi \left( A_{{x}}{+A}_{{y}}\right) +\sin \chi \left( {B}%
_{{x}}{+B}_{{y}}\right) , \\
\zeta _{Y} &=&-\sin \chi \left( A_{{x}}{+A}_{{y}}\right) +\cos \chi \left( {B%
}_{{x}}{+B}_{{y}}\right) ,
\end{eqnarray*}%
where ${A}_{{x}}=-\varepsilon _{x^{\prime }}\cos \phi /\lambda _{+}$, $%
A_{y}=\varepsilon _{y^{\prime }}\sin \phi /\lambda _{-}$, ${B}_{{x}%
}=\varepsilon _{x^{\prime }}\cos \theta \sin \phi /\lambda _{+}$, $%
B_{y}=\varepsilon _{y^{\prime }}\cos \theta \cos \phi /\lambda _{-}$. The
condition $\left\langle \delta B_{X}\delta B_{Y}(t)\right\rangle =0$ simply
means that $\left\langle \boldsymbol{\zeta }_{X}\boldsymbol{\zeta }%
_{Y}\left( t\right) \right\rangle $ $=0$. Substituting each component of $%
\boldsymbol{\zeta }$ into the equation and considering that $\left\langle
\varepsilon _{i}\varepsilon _{j}\left( t\right) \right\rangle =\delta
_{ij}\sigma _{\varepsilon }^{2}f_{c}\left( \left\vert t\right\vert /\tau
_{c}\right) $, the Euler angle $\chi $ can be determined as
\begin{equation}
\tan 2\chi =\frac{2\left( \lambda _{+}^{2}-\lambda _{-}^{2}\right)
n_{x^{\prime }}n_{y^{\prime }}n_{z}}{\lambda _{+}^{2}\left( n_{y^{\prime
}}^{2}-n_{z}^{2}n_{x^{\prime }}^{2}\right) +\lambda _{-}^{2}\left(
n_{x^{\prime }}^{2}-n_{z}^{2}n_{y^{\prime }}^{2}\right) },
\end{equation}%
where $\boldsymbol{\hat{n}}$ is the direction of the magnetic field.

With the knowledge of all the Euler angles, we can now calculate $%
\left\langle \zeta _{X}\zeta _{X}\left( t\right) \right\rangle $ and $%
\left\langle \zeta _{Y}\zeta _{Y}\left( t\right) \right\rangle $,
\begin{eqnarray*}
\left\langle \zeta _{X}\zeta _{X}\left( t\right) \right\rangle &=&\frac{1}{%
\Lambda _{+}^{2}}\sigma _{\varepsilon }^{2}f_{c}\left( \left\vert
t\right\vert /\tau _{c}\right) , \\
\left\langle \zeta _{Y}\zeta _{Y}\left( t\right) \right\rangle &=&\frac{1}{%
\Lambda _{-}^{2}}\sigma _{\varepsilon }^{2}f_{c}\left( \left\vert
t\right\vert /\tau _{c}\right) ,
\end{eqnarray*}%
where the effective SO length is given by%
\begin{equation*}
\frac{2}{\Lambda _{\pm }^{2}}=\frac{1-n_{x^{\prime }}^{2}}{\lambda _{-}^{2}}+%
\frac{1-n_{y^{\prime }}^{2}}{\lambda _{+}^{2}}\pm \sqrt{\left( \frac{%
1-n_{x^{\prime }}^{2}}{\lambda _{-}^{2}}+\frac{1-n_{y^{\prime }}^{2}}{%
\lambda _{+}^{2}}\right) ^{2}-\frac{4n_{z^{\prime }}^{2}}{\lambda
_{+}^{2}\lambda _{-}^{2}}}.
\end{equation*}%
The magnetic correlators are thus
\begin{eqnarray}
J_{XX}(t) &=&\left[ \frac{2eB\sigma _{\varepsilon }}{\Lambda _{+}m^{\ast
}\omega _{d}^{2}}\right] ^{2}f_{c}\left( \left\vert t\right\vert /\tau
_{c}\right) ,  \label{Jxxt} \\
J_{YY}(t) &=&\left[ \frac{2eB\sigma _{\varepsilon }}{\Lambda _{-}m^{\ast
}\omega _{d}^{2}}\right] ^{2}f_{c}\left( \left\vert t\right\vert /\tau
_{c}\right) ,  \label{Jyyt}
\end{eqnarray}%
and $J_{ZZ}(t)=0$, as mentioned earlier. In the following discussion, we
choose the cutoff function $f_{c}\left( \left\vert t\right\vert /\tau
_{c}\right) $ to be exponential for simplicity (see Appendix \ref{append::cutoff functions} for other
types of cutoff functions).

\section{Spin Relaxation}

\begin{figure}[th]
% Requires \usepackage{graphicx}
\includegraphics[scale=0.35]{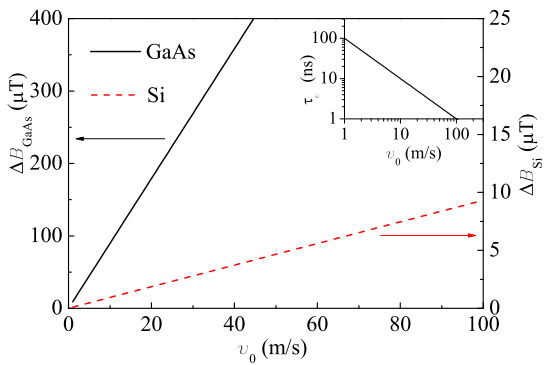}
\caption{$\Delta {B}$ as a function of moving velocity for GaAs (solid line)
QD with $\beta =300 $ m/s and Si (dashed line) QD with $%
\alpha =5$ m/s. The inset gives the bath correlation time $\tau _{c}$%
.}
\label{Fig2}
\end{figure}

Now we study decoherence of the electron spin qubit due to Hamiltonian (\ref%
{Heff}). The noise correlation time $\tau _{c}$ is generally much shorter
than the qubit decay time (the inset of Fig.~\ref{Fig2} shows values of $%
\tau _{c}$). In this regime, the dynamics of the spin qubit are governed by
the Bloch equations~\cite{Slichter1980}. %
%The general solutions are $\langle S_{X}(t)\rangle =S_{\perp }e^{-t/T_{2}}\sin
%(\omega t+\phi )$, $\langle S_{Y}(t)\rangle =S_{\perp }e^{-t/T_{2}}\cos
%(\omega t+\phi )$, and $\langle S_{Z}(t)\rangle =S_{T}+\left(
%S_{Z}^{0}-S_{T}\right) e^{-t/T_{1}}$, with the initial value $\langle
%\boldsymbol{S}(0)\rangle =(S_{\perp }\sin \phi ,S_{\perp }\cos \phi
%,S_{Z}^{0})$. At the low temperature limit ($T\ll \hbar \omega _{0}$) and
With purely transverse fluctuations, the longitudinal and transverse
relaxation rates, $1/T_{1}$ and $1/T_{2}$, are \cite{Golovach2004,
Borhani2006, Slichter1980}
\begin{eqnarray}
\frac{1}{T_{1}} &=&\frac{2}{T_{2}}=J_{XX}^{+}(\omega _{Z})+J_{YY}^{+}(\omega
_{Z}), \\
J_{ij}^{+}(\omega ) &=&\frac{g^{2}\mu _{B}^{2}}{2\hbar ^{2}}\int_{-\infty
}^{+\infty }\left\langle \delta B_{i}(0)\delta B_{j}(t)\right\rangle \cos
\left( \omega t\right) dt\,.  \notag
\end{eqnarray}%
Using Eq.~(\ref{Jxx}) and its $J_{YY}$ correspondent, we obtain
\begin{eqnarray}
\frac{1}{T_{1}}=\left[ \frac{2e\sigma _{\varepsilon }}{\hbar \omega _{d}^{2}}%
\right] ^{2}\frac{\omega _{Z}^{2}\tau _{c}}{1+\omega _{Z}^{2}\tau _{c}^{2}}%
F_{SO}(\theta ,\phi ),\hspace*{0.93in} &&  \label{1T1} \\
F_{SO}=\left[ \left( \beta ^{2}+\alpha ^{2}\right) \left( 1+\cos ^{2}\theta
\right) +2\alpha \beta \sin ^{2}\theta \cos 2\phi \right] . &&  \label{F_SO}
\end{eqnarray}%
Here $\theta $ and $\phi $ are the polar and azimuthal angles of $%
\boldsymbol{B}$ in the $x^{\prime }y^{\prime }z^{\prime }$ coordinates.

Before delving into the numerics we first discuss some qualitative features
of the spin relaxation rate here. Firstly, $1/T_{1}\propto 1/\omega _{d}^{4}$%
. This strong dependence on the QD confinement means that this spin
relaxation channel can be suppressed by having strong confinement for the
QD. Secondly, $1/T_{1}\propto \sigma _{\varepsilon }^{2}$. The origin of the
static disorder would determine the magnitude here. For example, in a
modulation doped GaAs structure, $\delta V\sim 20$ mV \cite{Nixon1990} and $%
l_{\varepsilon }\sim 0.1\mu $m \cite{Nixon1990}, so that $%
\sigma_{\varepsilon }=\delta V/l_{\varepsilon }\sim 200$ kV/m. On the other
hand, for an undoped top-gate structure in Si \cite{Maune2012}, there could
be disorder from defects in the barrier, though its characteristic length
and strength are unknown (most probably much smaller than in the modulation
doped structures). Our numerical estimates below use parameters from the
modulation doped structures.

The SO coupling dependence of $1/T_{1}$ is contained in $F_{SO}$ in terms of
$\alpha $ and $\beta $, the Rashba and Dresselhaus SO coupling strength.
These parameters are materials- and device-specific. In Si $\beta =0$, while
in GaAs $\beta _{GaAs}=300$ m/s is fixed (see Appendix \ref{append::SAW}). In both materials
$\alpha $ is dependent on the particular quantum well structure and doping.
%
%With these simplifications, the last bracket of Eq.~(\ref{1T1_v2}) becomes $\beta^2 (1+\cos^2 \theta)$ for GaAs, and $\alpha^2 (1+\cos^2 \theta)$ for Si.
%

The dependence on the direction of magnetic field $\boldsymbol{B}$ by $%
1/T_{1}$ is also contained in $F_{SO}$, in terms of the polar and azimuthal
angles $\theta $ and $\phi $. %
%The direction of the $\boldsymbol{B}$-field is given by the polar angle $\theta $
%between the field and the $z^{\prime }$ axis, and its azimuthal angle $\phi $
%(the in-plane rotation around the $z^{\prime }$ axis).
%
For example, for a perpendicular field ($\boldsymbol{B}\parallel \lbrack
001] $), $F_{SO}=2\left( \beta ^{2}+\alpha ^{2}\right) $. For an in-plane
field, $F_{SO}=\beta ^{2}+\alpha ^{2}+2\alpha \beta \cos 2\phi $. Thus, the
decay rate in a perpendicular field is always larger than if the field is
in-plane $\left( 1/T_{1}\right) _{perp}\geq \left( 1/T_{1}\right)
_{in-plane} $.

In the case of an in-plane magnetic field, the spin relaxation rate $1/T_{1}$
has a sinusoidal dependence on the azimuthal angle $\phi $ of the $%
\boldsymbol{B}$ field.
%Note that Eq.~(\ref{1T1//}) describes the distance of two vectors $\vec{a}$ and $\vec{b}$ with the magnitudes being proportional to $\left\vert \alpha \right\vert $ and $\left\vert \beta \right\vert $ and the
%angle $\vartheta $ between them being $\pi -2\phi $, and
The minimum rate is $1/T_{1}=\left[ 2e\sigma _{\varepsilon }\left( \beta
-\alpha \right) /\left( \hbar \omega _{d}^{2}\right) \right] ^{2}\omega
_{Z}^{2}\tau _{c}/\left( 1+\omega _{Z}^{2}\tau _{c}^{2}\right) $ (assuming $%
\alpha \beta >0$), when the $\boldsymbol{B}$ field is along the $y$ axis ($%
\phi =\pi /2$). Thus, in the special case when $\alpha =\beta $ and $\phi
=\pi /2$ (or $\alpha =-\beta $ and $\phi =0$), $1/T_{1}=0$. In other words,
since $\Delta {\boldsymbol{B}}$ is along the $y$ ($x$) axis when $\alpha =\beta $
($\alpha =-\beta $) [c.f. Eq. (\ref{DeltaB})], spin relaxation due to QD
motion \emph{vanishes} if the applied magnetic field $\boldsymbol{B}_{0}$ is
along $y$ for $\alpha =\beta $ (or along the $x$ axis for $\alpha =-\beta $%
). {Such special cases ($\alpha =\pm \beta $) have been discussed previously
in the context of spin relaxation due to phonon emission \cite%
{Schliemann2003, Golovach2004}.} Note that Hamiltonian $H$ in Eq. (\ref{H})
conserves the spin component $\sigma _{y(x)}$ for $\alpha =\beta $ ($\alpha
=-\beta $) and $\boldsymbol{B}_{0}\parallel y\,(x)$. This spin conservation
results in $T_{1}$ being infinite to all orders in the SO interaction
Hamiltonian (\ref{Hsop}). Meanwhile, decoherence rate $1/T_{2}$ reduces to
the next order contribution of Eq. (\ref{Hso}), in the form of pure dephasing.

The dependence on the magnitude of $\boldsymbol{B}$ and velocity $%
\boldsymbol{v}_{0}$ by $1/T_{1}$ is contained in $\omega
_{Z}^{2}\tau_{c}/\left( 1+\omega_{Z}^{2}\tau _{c}^{2}\right) $ of Eq.~(\ref%
{1T1}).
%Before we discuss the relaxation rate dependence on $B$ and $\boldsymbol{v}_{0}$,
Here we first estimate the magnitude of $\Delta B$ [c.f. Eq. (\ref{DeltaB}%
)]. In Fig.~\ref{Fig2}, we plot $\Delta B$ as a function of velocity $|%
\boldsymbol{v}_{0}|$ for GaAs and Si QDs. $\Delta B$ is two orders of
magnitude larger in GaAs than in Si, but still negligible if a strong
magnetic field (order of Tesla) is applied. We discuss the high and low
field cases separately as follows.

%\textit{High field and slow moving limit---}

\subsection{High field and slow moving limit}

For a strong applied magnetic field ($B\geq 1$ T) and a slow moving QD (1
nm/ns $<v_{0}<$ 100nm/ns), $\Delta {\boldsymbol{B}}$ can be neglected, and
the condition $\omega_{Z}\tau _{c}\gg 1$ is satisfied. In this limit, the $%
\omega _{Z}$ (or $B$) dependence cancels out in Eq.~(\ref{1T1}). $%
1/T_{1} $ depends linearly on the speed $v_{0}$ of the QD, and is
independent of the direction of the motion (since $\Delta B$ is neglected).

\begin{figure}[t]
% Requires \usepackage{graphicx}
\includegraphics[scale=0.4]{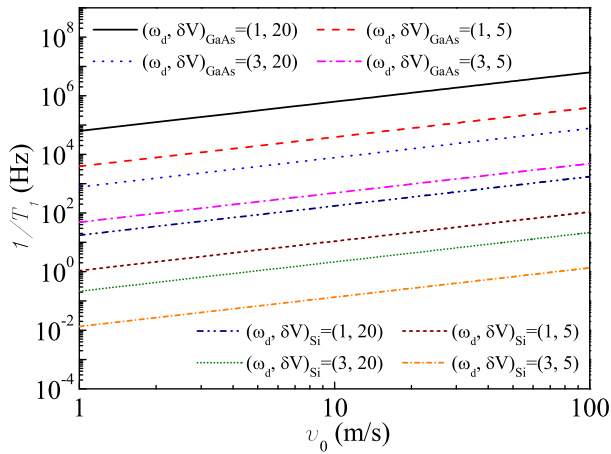}
\caption{Spin relaxation rate $1/T_{1}$ as a function of the QD velocity for
GaAs QDs with $\beta =300$ m/s and Si QDs with $\alpha =5$
m/s (with an in-plane magnetic field). Here $\omega _{d}$ and ${%
\delta }V$ are in units of meV and mV.}
\label{Fig3}
\end{figure}

We carry out numerical calculations on two representative QD structures, one
in GaAs/Al$_{1-x}$Ga$_{x}$As, the other in Si/SiGe. In both cases, the dot
confinement energies are set at $\hbar \omega _{d}=1$ meV and 3 meV,
%($\omega _{0}=1.5\times 10^{12}Hz$)
and the applied magnetic field is $B_{0}=1$ T. For the GaAs QD, we use the
bulk g factor $g=-0.44$, and the electron effective mass $m^{\ast
}=0.067m_{0}$, where $m_{0}$ is the free electron rest mass. For the Si QD, $%
g=2$, $m^{\ast }=0.19m_{0}$, and the Rashba SO coupling strength is chosen
to be $\alpha _{Si}=5$ m/s \cite{Wilamowski2002,Tahan2005,Prada2011}. %
%The correlation time $\tau _{c}$ of the fluctuating magnetic field $\delta B$ (as shown in the inset of Fig. 2)  And the limit $\omega \tau _{c}\gg 1$ is satisfied for $B_{0}>1T$, we thus
%obtain the linear dependence of $1/T_{1}$ with the velocity $1/T_{1}=\frac{2 \bar{H}_{\bot }^{2}}{\omega ^{2}\tau _{c}}\propto v/l_{0}$ (as shown in Fig.
%3).
%
%
Figure~\ref{Fig3} shows the spin relaxation rate $1/T_{1}$ as a function of
the QD speed in an in-plane $\boldsymbol{B}$ field, when
\begin{equation}
\left. \frac{1}{T_{1}}\right\vert _{\mathrm{in-plane}}=\frac{v_{0}}{%
l_{\varepsilon }}\left[ \frac{2e\sigma _{\varepsilon }}{\hbar \omega _{d}^{2}%
}\right] ^{2}\left( \beta ^{2}+\alpha ^{2}+2\alpha \beta \cos 2\phi \right)
\,.  \label{1T1//}
\end{equation}%
For a moving GaAs QD, we find that $T_{1}$ ranges from $\mu $s to 10 ms.
%, which is much shorter than $\tau =1$ ms known as the lower limit requirement of decoherence time for quantum computation \cite{tahan2002}.
For a Si QD, $T_{1}>1$ ms because of the weaker SO coupling. In terms of the
moving distance $\left( v_{0}T_{1}\right) _{in-plane}$,
%=\frac{l_{0}}{2}\left[ \frac{\hbar \omega _{0}^{2}}{e\sigma _{\varepsilon }}\right] ^{2}/\left( \beta^{2}+\alpha ^{2}+2\alpha \beta \cos 2\phi \right) $,
spin coherence is lost in as short as $\mu $m in GaAs and mm in Si for a dot
speed of 10 nm/ns.
%For the Si case, if it is not modulation doped, but with disorder, the random potential from the disorder is usually small, and this moving induced decay is %also small.
Clearly, while the spin relaxation times here are still much longer than
those in a 2DEG, the QD motion does present a serious threat to the
coherence of the spin qubit, especially in modulation doped GaAs
heterostructures. %
%And error correction technique may be needed \cite{Sugimoto2012}.
%The Zeeman frequency $\omega$ is eliminated, which means that the decay rate does not depends on the magnitude of magnetic field in our weak
%magnetic field assumption ($\omega <<\omega _{0}$). While, for dot confinement energy $\omega _{0}$, we have $1/T_{1}\varpropto \omega_{0}^{-4} $, which means %three times larger in confinement energy will result in two orders of reduction in decay rate. Furthermore, only the magnitude of the dot velocity affect the %decay rate, the direction of dot velocity does not contribute since $\Delta {\boldsymbol{B}}$ in Eq.~(\ref{DeltaB}) is negligible small.

%\textit{Low field and fast moving limit---}

\subsection{Low field and fast moving limit}

If the magnetic field is low, and/or the QD motion is fast (but still
adiabatic), so that $\omega _{Z}\tau _{c}\ll 1$, the spin relaxation rate is
\begin{equation}
\frac{1}{T_{1}}=\left[ \frac{2e\sigma _{\varepsilon }}{\hbar \omega _{d}^{2}}%
\right] ^{2}F_{SO}(\theta ,\phi )\omega _{Z}^{2}\tau _{c}.
\end{equation}%
If the applied field $B_{0}$ is much larger than $\Delta B$, so that $g\mu
_{B}\Delta B/\hbar \ll \omega _{Z}\ll \tau _{c}^{-1} $, we obtain $%
1/T_{1}\varpropto 1/v_{0}$, indicating motional narrowing. Whereas, if $%
B_{0}=0$, only $\Delta\boldsymbol{B}$ contributes to the spin splitting,
with $\theta =\pi /2$ and $\beta _{-}\tan \phi =\beta _{+}\cot \phi _{v}$.
Now $\omega _{Z}^{2}\tau _{c}=\left( 2m^{\ast }/\hbar \right)
^{2}v_{0}l_{\varepsilon }F_{v}\left( \phi _{v}\right) $, where $F_{v}\left(
\phi _{v}\right) =\left( \beta ^{2}+\alpha ^{2}+2\alpha \beta \cos 2\phi
_{v}\right) $. %
%Since $\tan \phi =\Delta B_{y}/\Delta B_{x}=\beta _{+}\cot \phi _{v}/\beta _{-}$,
%Eq. (\ref{F_SO}) is simplified as $F_{SO}^{2}(\pi /2,\phi _{v})=\left[ \beta
%^{2}+\alpha ^{2}+2\alpha \beta \left( \beta _{-}^{2}\sin ^{2}\phi _{v}-\beta
%_{+}^{2}\cos ^{2}\phi _{v}\right) /F_{v}\left( \phi _{v}\right) \right] $,
%
Interestingly, this $\phi _{v}$ dependence is completely canceled by that in
$F_{SO}$, so that
\begin{equation}
\frac{1}{T_{1}}=v_{0}l_{\varepsilon }\left[4e\sigma _{\varepsilon }m^{\ast
}\left( \beta ^{2}-\alpha ^{2}\right) /\left( \hbar \omega _{d}\right) ^{2}%
\right] ^{2}.
\end{equation}%
In other words, when no magnetic field is applied, $1/T_{1}$ depends
linearly on the speed $v_{0}$ of the QD motion and is independent of the
direction $\phi _{v}$ of the motion.

As an example we consider an SAW-confined electron in GaAs. Here the QD
moves fast, at the speed of sound $v_{0}=v_{SAW}=3000$ m/s, so that the
low-field limit $\omega _{Z}\tau _{c}\ll 1$ is satisfied even for $%
B_{0}\simeq 1$ T. The electron should still remain in the ground orbital
state, however, while the motion-induced magnetic field is now $\Delta {%
\boldsymbol{B}}\sim 0.02$ T. {The spin relaxation rate should have a weak
dependence on the direction of the motion when $B_{0}\ $and $\Delta B$ are
comparable.} The confinement potential for an electron in an SAW is
dependent on the driving intensity $P_{SAW}$, with $\omega _{d}\sim 1$ meV
(see Appendix \ref{append::SAW}).
%$\left( 1/T_{1}\right) _{in-plane}\varpropto 1/P_{SAW}^{2}$
%
%We estimate that the corresponding $\omega _{d}$ is about 1 meV .
%can be estimated as $\omega _{0}\approx \sqrt{E_{SAW}/m_{e}}%
%\omega _{SAW}/v_{SAW}$, where the energy $E_{SAW}$ characterize the
%vibration intensity/amplitude of SAW, which is proportional to the RF
%driving power $P$ \cite{Hermelin2011}, $\omega _{SAW}$ and $v_{SAW}$ are the
%frequency and velocity of SAW. Thus, $\left( 1/T_{1}\right)
%_{in-plane}\varpropto 1/E_{SAW}^{2}$. For $E_{SAW}\sim 80$ meV ($P\sim 20$
%dBm) and $\omega _{SAW}\sim 10$ GHz, the confinement potential is evaluated
%as $\omega _{0}\sim 1$ meV.
%
Using parameters for modulation doped samples, $l_{\varepsilon }\sim 0.1\mu $%
m and $\sigma _{\varepsilon }\sim 200$ kV/m, we estimate that $1/T_{1}\sim
10^{8}$ Hz in a strong in-plane magnetic field ($B_{0}\gg \Delta B$, $\beta
=300$ m/s and $\alpha =0$) and $1/T_{1}\sim 10^{5}$ Hz when $B_{0}=0$. These
rates can be reduced by having a larger $\omega _{d}$ (with higher SAW
driving intensity and/or higher frequency), and most importantly, using a
less disordered sample with smaller $\sigma _{\varepsilon }$.

\section{Discussion and Conclusion}

It is important to emphasize here that the static disorder potential is not the reservoir by itself.  It is the external driver that energizes the reservoir.  The disorder is what allows the spin to exchange energy with the reservoir: It modifies the mode functions of the electromagnetic environment.  As we mentioned at the beginning of the paper, we assume that the driving agent is classical and Markovian, so that
whatever information goes into the driving agent is lost. In other
words, we have performed a phenomenological study of an open system where
energy is not conserved. This calculation does not treat the external driving agent in a quantum mechanical fashion: It only energizes the disorder through motion, but does not have any internal structure. A definitive and complete study of the
spin-reservoir exchange requires quantization of the nonequilibrium
external driving agent, which is beyond the scope of the current study.
However, we hope this work could act as a catalyst for further theoretical
and experimental studies on this interesting problem.

%The expression of ${\Delta }B$ in Eq. (\ref{DeltaB}) and ${\delta }B$ in Eq.
%(\ref{dB}) show that if the dot velocity and the applied magnetic field are
%reversed, ${\Delta }B$ and ${\delta }B$ are reversed, too, so that at the
%lowest order of the SO coupling and Zeeman term, spin information can be
%restored by moving the electron backward with exactly the same path and
%speed under an exactly reversed applied magnetic field. However, if
%operations are carried out before the spin is moved back, the channel/path
%information would be lost, and it is no longer possible to restore the spin
%by just reversing its velocity and the applied magnetic field. More
%importantly, at higher orders, the effective field does not reverse exactly
%even if the reversing of the motion and the applied field is exact. This is
%clearly illustrated by the higher-order term ${\Delta }B_{L}$ in Eq. (\ref%
%{DeltaB_L}), which remains unchanged when the motion velocity and the
%applied field are reversed. Therefore, restoring spin information by
%reversing the path of the confined electron is practically unattainable.

As we mentioned before, for confined electron spins HF-induced dephasing is
the most important decoherence mechanism. For a moving but confined electron
spin, however, this dephasing is suppressed due to motional narrowing: $%
1/T_{2}^{\ast }\propto 1/v$ (see Appendix \ref{append::nuclear bath}). For example, in a QD with a 1
meV confinement energy in GaAs, the HF-induced dephasing time goes up from
30 ns in a stationary QD to $>1$ $\mu $s when the dot speed is 20 m/s, while
at this speed the spin relaxation due to SO coupling is already faster: $%
T_{1}\sim 1$ $\mu $s. Thus even with only a moderate speed of motion, the
HF-induced dephasing is already superseded by the SO-induced relaxation as
the main source of decoherence for the electron spin.

The model calculation presented in this manuscript focuses on the effect of the momentum scattering of the electron due to the random electrical potential in the plane of the quantum dot motion.  In our calculation we did not include the possible effect of a random component of the Rashba spin-orbit coupling due to the random electric field along the growth direction.\cite{Dugaev2009,Glazov2010}  This choice is a reflection of our assumption that the growth direction confinement is much stronger than the in-plane confinement, so that the relative fluctuation of the spin-orbit coupling should be weak compared to the existing coupling itself.  If this assumption does not hold, the random field along the growth direction would have to be properly taken into account as well.

In conclusion, we have studied electron spin relaxation in a moving QD. The
relaxation mechanism we studied originates from momentum scattering and SO
interaction. At the lowest order, it is a longitudinal relaxation channel,
so that $T_{2}=2T_{1}$. The relaxation rate is inversely proportional to the
fourth power of the confinement energy, so that spin decoherence is faster for
larger quantum dots. For high-field slow motion or very-low-field fast
motion, the decoherence rate increases linearly with the QD speed.
Quantitatively, in modulation-doped GaAs heterostructures this can be an
important spin decoherence channel, where spin relaxation time can be as
short as sub-$\mu $s and as long as ms, depending on the QD confinement
strength and the magnitude of the random potential. For modulation doped
Si/SiGe QDs the spin relaxation rate is generally several orders of
magnitude slower. However, compared to known spin decoherence channels in
Si, this relaxation can still be quite significant.

\section{acknowledgement}

We thank support by US ARO (W911NF0910393), DARPA QuEST through AFOSR, and
NSF PIF (PHY-1104672), and useful discussions with Seigo Tarucha and Eugene
Sherman.

%%%%%%%%%%%%%%%%
%
%   figures
%
%%%%%%%%%%%%%%%%

%\section*{Figures Captions}

%\bibliographystyle{apsrev}
%\bibliographystyle{plain}
%\bibliography{spinqubit}

\appendix

\section{Effects of the different cutoff functions}
\label{append::cutoff functions}

The decoherence of the moving electron spin $\boldsymbol{S}={\boldsymbol{%
\sigma }}/2$ is governed by the Hamiltonian (\ref{Heff}). In general, the
noise correlation time $\tau _{c}$ is much shorter than the spin decay time.
In this regime, the dynamics and relaxation of the spin is governed by the
Bloch equation~\cite{Slichter1980}. With purely transverse fluctuations, the
longitudinal and transverse relaxation rates, $1/T_{1}$ and $1/T_{2}$, are
\cite{Borhani2006, Slichter1980}
\begin{equation}
\frac{1}{T_{1}}=\frac{2}{T_{2}}=J_{XX}^{+}(\omega _{Z})+J_{YY}^{+}(\omega
_{Z}),
\end{equation}%
where the magnetic correlation function in the frequency domain is $%
J_{ij}^{+}(\omega )=\frac{g^{2}\mu _{B}^{2}}{2\hbar ^{2}}\int_{-\infty
}^{+\infty }\left\langle \delta B_{i}(0)\delta B_{j}(t)\right\rangle \cos
\left( \omega t\right) dt$. Thus
\begin{equation}
J_{XX}^{+}(\omega )=\frac{2\left( \omega _{Z}e\sigma _{\varepsilon }\right)
^{2}}{\left( \Lambda _{+}m^{\ast }\omega _{d}^{2}\right) ^{2}}\int_{-\infty
}^{+\infty }f_{c}\left( \left\vert t\right\vert /\tau _{c}\right) \cos
\left( \omega t\right) dt,  \label{Jxx}
\end{equation}%
and $J_{YY}^{+}(\omega )$ is obtained from Eq.~(\ref{Jxx}) by substituting $%
\Lambda _{+}\rightarrow \Lambda _{-}$. The relaxation rate is then
\begin{equation}
\frac{1}{T_{1}}=2\left[ \frac{\omega _{Z}e\sigma _{\varepsilon }}{\hbar
\omega _{d}^{2}}\right] ^{2}F_{SO}(\theta ,\phi )F_{c}\left( \omega
_{Z}\right) ,  \label{1T1_app}
\end{equation}%
where $F_{c}\left( \omega \right) $ is the Fourier transform of $f_{c}\left(
\left\vert t\right\vert /\tau _{c}\right) $, as shown in Table I.

\begin{table}[htb]
\centering
\begin{tabular}{|c|c|c|}
\hline
$F_{c}\left( \omega \right) $ & $\omega \tau _{c}\gg 1$ & $\omega \tau
_{c}\ll 1$ \\ \hline
$1$ & \multicolumn{1}{|r|}{$2/\left( \omega ^{2}\tau _{c}\right) $} &
\multicolumn{1}{|r|}{$2\tau _{c}$} \\ \hline
$2$ & \multicolumn{1}{|r|}{$\pi \tau _{c}\exp \left( -\left\vert \omega
\right\vert \tau _{c}\right) $} & \multicolumn{1}{|r|}{$\pi \tau _{c}$} \\
\hline
$3$ & \multicolumn{1}{|r|}{$\sqrt{\pi }\tau _{c}\exp \left( -\omega ^{2}\tau
_{c}^{2}/4\right) $} & \multicolumn{1}{|r|}{$\sqrt{\pi }\tau _{c}$} \\ \hline
\end{tabular}%
\caption{Approximations of $F_{c}(\omega )$ in different limits.}
\end{table}

\begin{figure}[htb]
% Requires \usepackage{graphicx}
\includegraphics[scale=0.4]{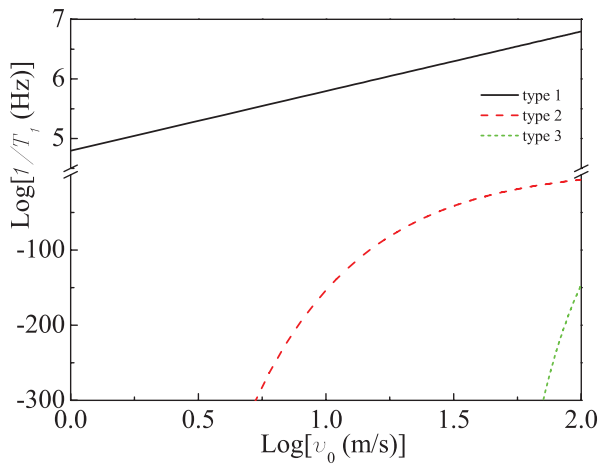}
\caption{Spin relaxation rate $1/T_{1}$ as a function of the velocity in
GaAs QDs for different types of correlation functions (in-plane field $B=1$ T, $\alpha =0$ and $\beta =300$ m/s).}
\label{FigTypes}
\end{figure}

As shown in Table II, different types of cutoff functions are very similar
at the low-field-fast-motion limit, but behave dramatically differently in
the high-field-slow-motion regime. We thus focus on the latter regime, and
plot the relaxation rate as a function of the QD velocity for different
types of correlations in Fig.~\ref{FigTypes}. Overall, $1/T_{1}$ is a
monotonically increasing function of the speed of the QD motion, no matter
which type of correlation function is used (in particular, $1/T_{1}$ is a
linear function of $v_{0}$ for the type-1 correlation function). However,
quantitatively the relaxation is completely suppressed for the type-2 and -3
cutoff functions because of the exponential suppression from $\exp \left(
-\left\vert \omega _{Z}\right\vert \tau _{c}\right) $ and $\exp \left(
-\omega _{Z}^{2}\tau _{c}^{2}/4\right) $ (with a 1 T external field in GaAs
and a $\tau _{c}$ between 1 and 100 ns, we are in the limit of $\omega
_{Z}\tau _{c}\gg 1$). In these cases, spin decoherence is probably dominated
by higher-order dephasing processes due to the SO coupling \cite{Makhlin2004}%
.

\section{Measuring spin-orbit coupling strength using carriers trapped by an
SAW}
\label{append::SAW}

A surface acoustic wave (SAW) in GaAs induces a piezo-electric field in the
form of $E_{SAW}\cos \left( k_{SAW}x-\omega _{SAW}t\right) $, where $%
k_{SAW}=2\pi /\lambda _{SAW}$ is the wave vector, and $\omega _{SAW}$ is the SAW
frequency. The troughs of this propagating electric potential can act as a
moving QD for electrons, with a velocity at the speed of sound. In the
moving frame the confinement potential is approximated as $E_{SAW}\cos
\left( k_{SAW}x\right) \approx E_{SAW}(1-k_{SAW}^{2}x^{2}/2)$ assuming $%
k_{SAW}x\ll 1$. Therefore, the confinement energy can be estimated as $%
\omega _{d}=2\pi \sqrt{E_{SAW}/m_{e}^{\ast }}/\lambda _{SAW}\varpropto \sqrt{%
P_{SAW}}/\lambda _{SAW}$, where $m_{e}^{\ast }=0.067m_{e}$ is the electron
effective mass, and $P_{SAW}$\ is the RF power that generates the SAW.
According to Ref. \onlinecite{Hermelin2011}, $E_{SAW}[eV]=2/25\times
10^{(P[dBm]-23)/20}$. We can thus estimate $E_{SAW}$ and $\omega _{d}$, as
shown in Table~\ref{table3}.

\begin{table}[h]
\centering
\begin{tabular}{|c|c|c|}
\hline
$P$ (dBm) & $E_{SAW}$ (meV) & $\omega _{d}$ (meV) \\ \hline
\multicolumn{1}{|r|}{$3$} & \multicolumn{1}{|r|}{$8$} & \multicolumn{1}{|r|}{%
$0.5$} \\ \hline
\multicolumn{1}{|r|}{$13$} & \multicolumn{1}{|r|}{$25.3$} &
\multicolumn{1}{|r|}{$0.9$} \\ \hline
\multicolumn{1}{|r|}{$23$} & \multicolumn{1}{|r|}{$80$} &
\multicolumn{1}{|r|}{$1.6$} \\ \hline
\end{tabular}%
\caption{Estimation of the confinement energies for different driving power.}
\label{table3}
\end{table}

SAW-trapped electrons can help determine the spin-orbit coupling strength in
the underlying material. For example, in GaAs the electrons and holes that
are photoexcited can be picked up by an SAW. The spatial distribution of
electron spins can then be measured by photoluminescence (specifically
polarization of the emitted photons) or magneto-optic Kerr rotation \cite%
{Stotz2005,Sanada2011}. In terms of Kerr rotation, for instance, the spin
distribution is expressed as $\theta _{K}(d)=\theta
_{0}e^{-d/v_{0}T_{2}}\cos (\omega _{Z}d/v_{0})=\theta _{0}e^{-d/L_{s}}\cos
(2\pi \kappa d)$, where $d$ is the distance from the origin where the
excitons are generated, $L_{s}=v_{0}T_{2}$ is the spin decay length, and $%
\kappa =\omega _{Z}/\left( 2\pi v_{0}\right) $ is the spatial precession
frequency. In the absence of an applied magnetic field, the spin precession
frequency $\omega _{Z}=g\mu _{B}\Delta B/\hbar $ is completely determined by
the motion-induced magnetic field $\Delta {\boldsymbol{B}}$,
\begin{equation}
\Delta {\boldsymbol{B}}=\frac{2m^{\ast }}{g\mu _{B}}\left( \beta
_{-}v_{0y},\beta _{+}v_{0x},0\right) \,.  \label{DeltaB_app}
\end{equation}%
This field has been measured to be sizable (25 mT in Ref.~%
\onlinecite{Stotz2005} for GaAs), because of the high speed of the QD
motion.
%which is large in the case of SAW because of its high speed as compared to the
%moving QDs induced by programming the surface gate potential. Here $\Delta B$ can be
%estimated at 26 mT for $v=3000$ m/s, $\beta =300$ m/s and $\alpha =0$ m/s.
Based on Eq. (\ref{DeltaB_app}) and $\kappa =\omega _{Z}/\left( 2\pi
v_{0}\right) =g\mu _{B}\Delta B/\left( hv_{0}\right) $, one can determine
the SO coupling constants by measuring the spatial precession frequency $%
\kappa $ experimentally \cite{Sanada2011}. This leads to an upper limit of
the Dresselhaus SO coupling constant at $\beta =300$ m/s, which is
consistent with other recent experiments \cite%
{Studer2010,Sanada2011,Zumbuhl2002}. However, this $\beta $ value is smaller
than what was used in earlier theoretical calculations \cite{Tahan2002,
Stano2006, Hanson2007}. More experimental studies would be needed to clarify
this issue.

\section{Motional Narrowing of Nuclear Spin Induced Dephasing}
\label{append::nuclear bath}

As we have discussed in the main text, the main decoherence channel for a
confined electron spin in a finite field is the hyperfine interaction
induced pure dephasing \cite{Cywinski2009, Bluhm2011}, while spin relaxation
of free electrons and holes in semiconductors is dominated by spin-orbit
interaction \cite{Zutic2004, Dugaev2009}. Here we show how the effect of
hyperfine interaction is strongly suppressed by motional narrowing for a
moving electron spin qubit with controlled motion.

\begin{figure}[htb]
% Requires \usepackage{graphicx}
\includegraphics[scale=0.35]{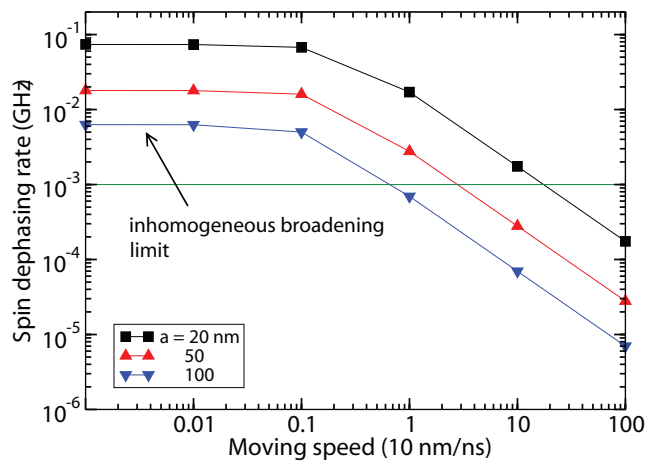}
\caption{Spin dephasing time $T_{2}^*$ as a function of the speed $v$ of a
moving GaAs QD. The horizontal line indicates a dephasing time of 1 $%
\mu$s.}
\label{FigOverhauser}
\end{figure}

The contact hyperfine interaction for the electron in a quantum dot can be
written as
\begin{equation}
H_{HF} = \sum A_i {\boldsymbol{S}} \cdot {\boldsymbol{I}}_i \,,
\end{equation}
where $A_i = A|\psi({\boldsymbol{r}}_i)|^2$ is the hyperfine coupling
constant at lattice site $i$ [with $A$ being the total hyperfine coupling
strength and $\psi({\boldsymbol{r}})$ the electron envelope function in the
quantum dot]. In GaAs, for example, $A = 92 \ \mu$eV, ${\boldsymbol{S}}$ is
the electron spin, and ${\boldsymbol{I}}_i$ is the nuclear spin at lattice
site $i$. Since nuclear spin evolves much more slowly compared to the
electron spin, we can treat it semiclassically as an Overhauser field:
\begin{equation}
{\boldsymbol{B}}_N = \sum A_i {\boldsymbol{I}}_i \,.
\end{equation}
In the presence of an external magnetic field along the $z$ direction, we can
focus on the $z$ component of the Overhauser field $B_{Nz}$ (let $g \mu_B =
1 $):
\begin{equation}
H = (B_0 + B_{Nz}) S_z.
\end{equation}
Since nuclear spins are randomly oriented in a sample at any reasonable
experimental temperature, the electron spin would experience a fluctuating
magnetic field as it moves, and undergo dephasing accordingly in the
off-diagonal element of the electron spin density matrix: $%
\rho_{\uparrow\downarrow} (t) = \rho_{\uparrow\downarrow} (0) e^{-\delta
\phi(t)}$. Here the phase diffusion is given by
\begin{equation}
\delta \phi(t) = \frac{1}{2\hbar^2} \int_0^{\infty} d\omega \
S_{B_N}(\omega) \left(\frac{\sin \omega t/2}{\omega/2}\right)^2.
\end{equation}
Here the nuclear field spectral density $S_{B_N}(\omega)$ is
\begin{equation*}
S_{B_N}(\omega) = \frac{1}{2\pi} \int_{-\infty}^{\infty} dt e^{i\omega t}
\langle B_{Nz}(t) B_{Nz}(0) \rangle.
\end{equation*}
For a moving quantum dot in GaAs with a trajectory ${\boldsymbol{r}}(t) = {%
\boldsymbol{r}}(0) + {\boldsymbol{v}} t$, and assuming that any two
different nuclear spins are completely uncorrelated, we find
%\begin{widetext}
\begin{equation*}
\langle B_{Nz}(t) B_{Nz}\rangle= \frac{5}{4} A^2 \Omega \int d{\boldsymbol{R}} \ |\psi({\boldsymbol{R}}
- {\boldsymbol{r}}(t))|^2 |\psi({\boldsymbol{R}} - {\boldsymbol{r}}(0))|^2,
\end{equation*}
%\begin{eqnarray*}
%\langle B_{Nz}(t) B_{Nz}(0) \rangle & = & \sum_{i,j} A^2 |\psi({\boldsymbol{R%
%}}_i - {\boldsymbol{r}}(t))|^2 |\psi({\boldsymbol{R}}_j - {\boldsymbol{r}}%
%(0))|^2 \langle I_{iz} I_{jz} \rangle = \frac{5}{4} A^2 \Omega \int d{\boldsymbol{R}} \ |\psi({\boldsymbol{R}}
%- {\boldsymbol{r}}(t))|^2 |\psi({\boldsymbol{R}} - {\boldsymbol{r}}(0))|^2
%\end{eqnarray*}
%\end{widetext}
where we have used $\langle I_z^2 \rangle = 5/4$ for GaAs, and $\Omega$ is
the volume of a lattice unit cell. For simplicity, we calculate the integral
using a Gaussian envelope wave function for the electron with radius $a$,
and obtain the spectral density as
\begin{equation}
S_{B_N}(\omega) = \frac{5A^2}{16\pi^2} \frac{\Omega}{a^3} \frac{a}{v}
e^{-\left(\frac{\omega a}{v}\right)^2/2} \,,
\end{equation}
where $a$ is the radius of the quantum dot, and $v$ is the speed of QD
motion. Now the electron spin phase diffusion can be calculated:
\begin{equation}
\delta \phi (t) = \frac{5 A^2}{16 \pi^2 \hbar^2} \frac{\Omega}{a^3} \frac{a t%
}{v} \int_0^\infty d\theta e^{-2(a/vt)^2 \theta^2} \left( \frac{\sin \theta}{%
\theta} \right)^2 \,.  \label{eq:dephasing_Overhauser}
\end{equation}
The integral here can be evaluated numerically given dot radius $a$ and dot
speed $v$. In Fig.~\ref{FigOverhauser} we plot the spin dephasing time $%
T_{2}^*$ as a function of the speed $v$ of a moving GaAs QD. Here $T_2^*$ is
defined according to the equality $\delta \phi(T_2^*) = 1$. When the motion
speed goes to zero, we obtain the inhomogeneous broadening for a fixed QD,
with dephasing time between 10 and 100 ns. When the speed $v$ is large ($%
>10$ m/s), the dephasing becomes suppressed. The electron samples a larger
number of nuclear spins as it moves faster, and the effect of the random
nuclear field averages out, which is a typical manifestation of the motional
narrowing effect. A close inspection of the high-speed results in Fig.~\ref%
{FigOverhauser} and Eq.~(\ref{eq:dephasing_Overhauser}) shows that for large
$v$, the dephasing time $T_2^*$ due to hyperfine interaction is proportional
to $v$, so the dephasing rate is proportional to $1/v$, while Eq.~(\ref{1T1}%
) shows that the spin relaxation rate due to spin-orbit interaction is
proportional to $v$. Comparing numerical results given in Figs. \ref{Fig3Gamma_v} and \ref{FigOverhauser}, we can see that in GaAs, when the dot
speed exceeds 10 m/s to 100 m/s, the nuclear spin induced dephasing becomes
less important than the spin-orbit and random potential induced relaxation.

%\bigskip

In the discussion here about nuclear spin induced dephasing, we have
considered only the lowest order effect of the nuclear spins, i.e., the
inhomogeneous broadening induced by random but static nuclear polarization.
We have not considered pure dephasing due to nuclear spin dynamics. How the
nuclear spin dynamics is affected by the dot motion, and how the dynamics
would feed back into the electron spin coherence/decoherence, remain open
theoretical questions. Qualitatively, nuclear spin dynamics induced electron
spin decoherence is generally slower than dephasing due to inhomogeneous
broadening. Thus the cross-over we study here should be a reliable
indication of the overall competition between hyperfine interaction and
spin-orbit interaction.

\end{document}